\documentclass[12pt]{article}
\usepackage{amssymb}
\usepackage{graphicx}
\begin{document}
\newcommand{\beq}{\begin{equation}}
\newcommand{\eeq}{\end{equation}}
\newcommand{\beqa}{\begin{eqnarray}}
\newcommand{\eeqa}{\end{eqnarray}}
\newcommand{\beqar}{\begin{eqnarray*}}
\newcommand{\eeqar}{\end{eqnarray*}}
\newcommand{\al}{\alpha}
\newcommand{\be}{\beta}
\newcommand{\del}{\delta}
\newcommand{\D}{\Delta}
\newcommand{\eps}{\epsilon}
\newcommand{\ga}{\gamma}
\newcommand{\Ga}{\Gamma}
\newcommand{\ka}{\kappa}
\newcommand{\nn}{\nonumber}
\newcommand{\inn}{\!\cdot\!}
\newcommand{\h}{\eta}
\newcommand{\ii}{\iota}
\newcommand{\kk}{\varphi}
\newcommand\F{{}_3F_2}
\newcommand{\la}{\lambda}
\newcommand{\La}{\Lambda}
\newcommand{\na}{\prt}
\newcommand{\Om}{\Omega}
\newcommand{\om}{\omega}
\newcommand{\p}{\Phi}
\newcommand{\sig}{\sigma}
\renewcommand{\t}{\theta}
\newcommand{\z}{\zeta}
\newcommand{\ssc}{\scriptscriptstyle}
\newcommand{\eg}{{\it e.g.,}\ }
\newcommand{\ie}{{\it i.e.,}\ }
\newcommand{\labell}[1]{\label{#1}} 
\newcommand{\reef}[1]{(\ref{#1})}
\newcommand\prt{\partial}
\newcommand\veps{\varepsilon}
\newcommand{\pol}{\varepsilon}
\newcommand\vp{\varphi}
\newcommand\ls{\ell_s}
\newcommand\cF{{\cal F}}
\newcommand\cA{{\cal A}}
\newcommand\cS{{\cal S}}
\newcommand\cT{{\cal T}}
\newcommand\cV{{\cal V}}
\newcommand\cL{{\cal L}}
\newcommand\cM{{\cal M}}
\newcommand\cN{{\cal N}}
\newcommand\cG{{\cal G}}
\newcommand\cK{{\cal K}}
\newcommand\cH{{\cal H}}
\newcommand\cI{{\cal I}}
\newcommand\cJ{{\cal J}}
\newcommand\cl{{\iota}}
\newcommand\cP{{\cal P}}
\newcommand\cQ{{\cal Q}}
\newcommand\cg{{\tilde {{\cal G}}}}
\newcommand\cR{{\cal R}}
\newcommand\cB{{\cal B}}
\newcommand\cO{{\cal O}}
\newcommand\tcO{{\tilde {{\cal O}}}}
\newcommand\bz{\bar{z}}
\newcommand\bb{\bar{b}}
\newcommand\bR{\bar{R}}
\newcommand\ba{\bar{a}}
\newcommand\bg{\bar{g}}
\newcommand\bc{\bar{c}}
\newcommand\bw{\bar{w}}
\newcommand\bX{\bar{X}}
\newcommand\bK{\bar{K}}
\newcommand\bA{\bar{A}}
\newcommand\bZ{\bar{Z}}
\newcommand\bxi{\bar{\xi}}
\newcommand\bphi{\bar{\phi}}
\newcommand\bpsi{\bar{\psi}}
\newcommand\bprt{\bar{\prt}}
\newcommand\bet{\bar{\eta}}
\newcommand\btau{\bar{\tau}}
\newcommand\hF{\hat{F}}
\newcommand\hA{\hat{A}}
\newcommand\hT{\hat{T}}
\newcommand\htau{\hat{\tau}}
\newcommand\hD{\hat{D}}
\newcommand\hf{\hat{f}}
\newcommand\hK{\hat{K}}
\newcommand\hg{\hat{g}}
\newcommand\hp{\hat{\Phi}}
\newcommand\hi{\hat{i}}
\newcommand\ha{\hat{a}}
\newcommand\hb{\hat{b}}
\newcommand\hQ{\hat{Q}}
\newcommand\hP{\hat{\Phi}}
\newcommand\hS{\hat{S}}
\newcommand\hX{\hat{X}}
\newcommand\tL{\tilde{\cal L}}
\newcommand\hL{\hat{\cal L}}
\newcommand\tG{{\tilde G}}
\newcommand\tg{{\tilde g}}
\newcommand\tphi{{\widetilde \Phi}}
\newcommand\tPhi{{\widetilde \Phi}}
\newcommand\te{{\tilde e}}
\newcommand\tk{{\tilde k}}
\newcommand\tf{{\tilde f}}
\newcommand\ta{{\tilde a}}
\newcommand\tb{{\tilde b}}
\newcommand\tc{{\tilde c}}
\newcommand\td{{\tilde d}}
\newcommand\tm{{\tilde m}}
\newcommand\tmu{{\tilde \mu}}
\newcommand\tnu{{\tilde \nu}}
\newcommand\talpha{{\tilde \alpha}}
\newcommand\tbeta{{\tilde \beta}}
\newcommand\trho{{\tilde \rho}}
 \newcommand\tR{{\tilde R}}
\newcommand\teta{{\tilde \eta}}
\newcommand\tF{{\widetilde F}}
\newcommand\tK{{\tilde K}}
\newcommand\tE{{\widetilde E}}
\newcommand\tpsi{{\tilde \psi}}
\newcommand\tX{{\widetilde X}}
\newcommand\tD{{\widetilde D}}
\newcommand\tO{{\widetilde O}}
\newcommand\tS{{\tilde S}}
\newcommand\tB{{\tilde B}}
\newcommand\tA{{\widetilde A}}
\newcommand\tT{{\widetilde T}}
\newcommand\tC{{\widetilde C}}
\newcommand\tV{{\widetilde V}}
\newcommand\thF{{\widetilde {\hat {F}}}}
\newcommand\Tr{{\rm Tr}}
\newcommand\tr{{\rm tr}}
\newcommand\STr{{\rm STr}}
\newcommand\hR{\hat{R}}
\newcommand\M[2]{M^{#1}{}_{#2}}
\newcommand\MZ{\mathbb{Z}}
\newcommand\MR{\mathbb{R}}
\newcommand\bS{\textbf{ S}}
\newcommand\bI{\textbf{ I}}
\newcommand\bJ{\textbf{ J}}

\begin{titlepage}
\begin{center}

\vskip 2 cm
{\LARGE \bf  
T-duality and background-dependence in  \\  \vskip 0.25 cm genus corrections to effective actions
 }\\
\vskip 1.25 cm
 Mohammad R. Garousi\footnote{garousi@um.ac.ir}

\vskip 1 cm
{{\it Department of Physics, Faculty of Science, Ferdowsi University of Mashhad\\}{\it P.O. Box 1436, Mashhad, Iran}\\}
\vskip .1 cm
 \end{center}

\begin{abstract}

      The classical effective action in string theory is background-independent, and its invariance under the Buscher rules constrains its form up to a few parameters. This work investigates how this picture changes at the quantum level, where loop corrections introduce an inherent background dependence.

    We propose a T-duality map for the loop-level effective action. It connects the circle-reduced effective action at large radius—where loops include only Kaluza-Klein (KK) momentum modes—to the base-space effective action at small radius—where loops include only winding modes. The resulting effective action is fundamentally distinct: it cannot be obtained from the KK reduction of any standard higher-dimensional action, revealing a uniquely stringy phenomenon at the loop level.

\end{abstract}

\end{titlepage}

\section{Introduction}

The spacetime effective action in string theory exhibits a double expansion: in the world-sheet genus $g$ and in the spacetime derivative parameter $\alpha'$. Several complementary approaches have been developed to determine the  $\alpha'$-expansion, including  the S-matrix method \cite{Gross:1986iv,Gross:1986mw}, the non-linear sigma model \cite{Tseytlin:1988rr}, T-duality \cite{Garousi:2017fbe}, and supersymmetry \cite{Ozkan:2024euj}.
The supersymmetry method, applicable exclusively to superstring theory, leverages spacetime supersymmetry to construct the effective action. In contrast, the non-linear sigma model and T-duality approaches rely on the conformal symmetry of the world-sheet—a universal symmetry inherent to all string theories. In this work, we focus on investigating higher-genus corrections through the lens of T-duality.

In the framework of the non-linear sigma model, the spacetime equations of motion for massless background fields at genus $g$ are obtained by imposing conformal invariance on the two-dimensional model up to that order. The beta functions, computed using two-dimensional field theory, are set to zero according to the condition \(\sum_{n=0}^g \beta_n = 0\). This condition yields the spacetime equations of motion and, consequently, determines the corresponding effective action \cite{Callan:1986bc}.
Summing beta functions across genus orders is necessary because integration over the Teichmüller space at a given genus often diverges when handles or boundaries degenerate. These divergences are canceled by introducing appropriate counterterms from lower genus levels \cite{Fischler:1986tb}.

The condition \(\sum_{n=0}^g \beta_n = 0\) implies that the non-linear sigma model at genus order \(g\) is not independently conformally invariant. Rather, the anomalies from lower-genus orders combine to cancel the anomaly at genus \(g\). In contrast, when the effective action is derived by enforcing spacetime symmetries, no such divergences arise. Consequently, the effective action at each genus order may be expected to exhibit independent invariance under the possible spacetime symmetries of string theory. Within this framework, however, contributions from lower-genus orders can appear as genus corrections to the symmetry transformations.

At a given genus order $g$, each beta function has its own $\alpha'$-expansion, corresponding to loop calculations in the two-dimensional field theory. Specifically, the beta function at order $\alpha'^m$ is associated with $(m+1)$-loop calculations.  In contrast, when deriving the effective action by imposing spacetime symmetries, such contributions may manifest as  derivative corrections to the symmetry transformations. The non-linear sigma model approach has been successfully employed at the sphere level ($\beta_0 = 0$) to derive gravity couplings up to order $\alpha'^3$ \cite{Grisaru:1986kw,Grisaru:1986vi} and at the torus level ($\beta_0 + \beta_1 = 0$) to compute the cosmological constant in bosonic string theory \cite{Fischler:1986tb}.

The conformal symmetry of the world-sheet theory requires that non-linear sigma models formulated in two spacetime backgrounds with circular isometries be related via Buscher transformations \cite{Buscher:1987sk,Rocek:1991ps}. Since these transformations are genus-independent \cite{Hamidi:1986vh}, the classical effective action of string theory—along with its higher-genus corrections—must remain invariant under Buscher transformations for any background admitting circular isometries. This invariance constrains the effective action in the critical dimension \( D \), where conformal symmetry is preserved.

At the classical level, the $D$-dimensional spacetime effective action at any order of \(\alpha'\) is background-independent \cite{Garousi:2022ovo}. This implies that the coupling constants for backgrounds with circular isometries are identical to those for arbitrary backgrounds. Consequently, the requirement of T-duality invariance in the $(D-1)$-dimensional base space effective action can be systematically utilized to determine these coupling constants in the $D$-dimensional effective action for any background.
To systematically implement this constraint on the $D$-dimensional classical effective action, one first identifies all independent covariant and gauge-invariant couplings at a given order in $\alpha'$, each parametrized by an undetermined coefficient. The requirement that the KK reduction of the classical effective action be T-duality invariant then fixes these coefficients in terms of a finite set of parameters at each order in $\alpha'$. Crucially, the standard Buscher rules must be applied to the two-derivative couplings, while their extensions—incorporating $\alpha'$-corrections—must govern the higher-derivative terms \cite{Kaloper:1997ux,Garousi:2019wgz}.
   At the sphere level, this approach has successfully derived NS-NS couplings up to order $\alpha'^3$ \cite{Kaloper:1997ux,Garousi:2019wgz,Garousi:2019mca,Garousi:2020gio}, with results consistent with both the non-linear sigma model \cite{Grisaru:1986kw,Grisaru:1986vi} and S-matrix calculations \cite{Gross:1986iv,Gross:1986mw}.  Hence, the background independence of the classical couplings allows one to systematically determine both the $D$-dimensional effective action and, in the case where one spacetime dimension is a circle, the $(D-1)$-dimensional base space effective action by using the KK reduction of the $D$-dimensional effective action.

In contrast to the classical effective action, quantum corrections are inherently background-dependent. Specifically, the loop-level couplings in a spacetime with a circular dimension differ fundamentally from those in a spacetime without one. This difference arises because, in the presence of a circle, the loop-level S-matrix elements—from which couplings are extracted—involve both KK momentum and winding momentum \cite{Green:1982sw}. The inclusion of winding momentum, a purely stringy phenomenon, results in couplings that are entirely distinct from those in a non-compact spacetime, where winding modes are absent.
This background dependence poses a significant challenge: T-duality cannot be easily imposed to determine the $(D-1)$-dimensional loop-level effective action for a spacetime with a circle. In fact, at the quantum level, T-duality does not constrain the $D$-dimensional effective action itself.

However, in this paper we speculate on a potential pathway. In the limit of a very large circle radius, the effects of winding momentum in loops become negligible. In this regime, the $(D-1)$-dimensional couplings might be derivable via KK reduction from the $D$-dimensional theory, which naturally lacks winding modes. The resulting couplings would be valid only in the large-radius limit and would not be T-duality invariant\footnote{The erroneous assumption of background independence for the loop-level effective action forces an incorrect T-duality constraint on its dimensionally reduced couplings, leading to the false conclusion that T-duality fails for genus corrections \cite{Garousi:2024pqc}.}. Applying a T-duality transformation to these couplings would then generate the correct couplings for the opposite corner of moduli space, where the circle radius is very small. These small-radius couplings, being pure stringy effects, cannot be obtained from any covariant, gauge-invariant $D$-dimensional action via standard KK reduction.

In this paper, we employ the aforementioned proposal to investigate loop-level effective actions. Section 2 begins with a review of the bosonic string's loop-level cosmological constant, emphasizing its background dependence. We demonstrate its invariance under the Buscher rules when one dimension is compactified on a circle. Furthermore, we elucidate how the $(D-1)$-dimensional partition function at a large radius is related to the $D$-dimensional one via KK reduction. We then show how applying T-duality to this large-radius result yields the $(D-1)$-dimensional partition function for a small radius—a purely stringy state unobtainable from the KK reduction of the $D$-dimensional theory.
Building on this foundation, Section 3 extends the T-duality analysis to the loop-level effective action of heterotic string theory. In particular, we show that T-duality requires the inclusion of a moduli-dependent function for the couplings. We reveal a key quantum effect: couplings excluded by T-duality at the classical level may, in fact, be permitted in the loop-level effective action.
Section 4 then presents a parallel T-duality analysis for the loop-level effective action of type II string theory. In Subsection 4.1, we show that the moduli functions in type II theories are constrained by S-duality to possess only large-radius and small-radius limits. Finally, Section 5 provides a brief discussion of our results.

\section{T-duality of cosmological constant}

At the quantum level, the leading contribution to the effective action is the zero-derivative term, the cosmological constant. For bosonic string theory at genus $g = 1$, the cosmological constant in Minkowski spacetime is non-zero and is given by the modular integral of the torus partition function $Z_1$ over the fundamental region $\mathcal{F}$ of the moduli space (see, e.g., \cite{Polchinski:1985zf,Peskin:1987rz}). Specifically, 
\beqa
\Lambda_1 \sim \int_{\cF} \frac{d^2\tau}{({\rm Im}\tau)^2} Z_1(\tau).
\eeqa
More generally, at the $g$-loop level, the cosmological constant $\Lambda_g$ for oriented closed strings is obtained as the modular integral of the partition function $Z_g$ over the fundamental region of the moduli space for a genus-$g$ world-sheet. The corresponding term in the effective action at the critical dimension $D$ takes the form:
\beqa
\int d^Dx \, e^{2(g-1)\Phi} \sqrt{-G} \, \Lambda_g.\labell{Dc}
\eeqa
Explicit calculations in string theory reveal that partition functions - and by extension, cosmological constants - exhibit a direct dependence on the spacetime background. A particularly instructive example arises when considering backgrounds with one compact circular dimension, where these dependencies become clearly manifest.
 
When one spatial dimension is compactified on a circle of  dimensionless radius \( R=e^{\vp/2 }\), the metric and dilaton are reduce as \cite{Maharana:1992my,Kaloper:1997ux}
\beqa  
G_{\alpha\beta} = \left(\matrix{\bg_{ab}  & 0 \cr 0 & e^\vp  &}\!\!\!\!\!\right)\,,\,\,\,\,\Phi =  \bar{\phi} + \varphi/4.\labell{red}
\eeqa  
We set the off-diagonal metric $G_{a y}$ to zero, as it does not appear in our discussion of the partition function.  In this case, there are two key modifications arise in the loop-level partition function calculation.  
 First, the integral over the momentum along this direction is replaced by a summation over KK momenta. Second, winding modes along this direction must also be taken into account. The contribution of the compact circle to the partition function (see, e.g., \cite{Giveon:1994fu,Alvarez:1994dn}) takes the form:  
\beqa  
Z_g(\vp,\tau) = Z_g'(\tau) e^{-g\vp/2} \det({\rm Im}\,\tau) \sum_{K, M} e^{-2\pi i{\rm Re}\,\tau KM - \pi {\rm Im}\,\tau (K^2e^{-\vp} + M^2 e^\vp)}, \labell{PF} 
\eeqa  
where \( Z_g' \) represents the contribution from non-compact directions.  
Incorporating the  dilaton factor \( e^{2(g-1)\Phi} \) and the \( e^{\vp/2} \) factor from the reduction of \( \sqrt{-G} =e^{\vp/2}\sqrt{-\bg}\), the full partition function becomes:  
\[  
Z_g(\vp,  \bar{\phi}, \tau) = e^{2(g-1) \bar{\phi}} e^{g\vp/2} Z_g(\vp, \tau).  
\]  
This leads to the following term in the $(D-1)$-dimensional effective action:  
\beqa  
2\pi\int d^{D-1}x \,  \sqrt{-\bg} \, \Lambda_g(\vp, \bar{\phi}),
\labell{cosm}  
\eeqa  
which explicitly demonstrates its dependence on the background fields. 
The $(D-1)$-dimensional effective action presented in \reef{cosm} is not derivable via standard KK reduction from the $D$-dimensional Minkowski action in \reef{Dc}. A key obstruction is the absence of winding momentum contributions in the loop amplitudes of the Minkowski action.

Under the Buscher transformations \cite{Buscher:1987sk,Rocek:1991ps}, 
\beqa  
\bg'_{\mu\nu}=\bg_{\mu\nu}, \quad  \bar{\phi}' = \bar{\phi}, \quad \vp' = -\vp, 
\labell{backg}  
\eeqa  
the $(D-1)$-dimensional effective action \reef{cosm} remains invariant. The measure is invariant; the cosmological constant is also invariant, i.e.,
\beqa
\Lambda_g(-\vp,\bar{\phi})&=&\Lambda_g(\vp,\bar{\phi})\,.
\eeqa
This invariance occurs because the partition function \reef{PF} includes both KK and winding momenta in the loop momentum for the circle direction.     The presence of both KK and winding momenta also implies that the partition function must be invariant under the modular transformations $\tau\rightarrow \tau+1$ and $\tau\rightarrow -1/\tau$ (see, e.g., \cite{Becker:2007zj}).

    In the limit of large $\vp$, the partition function \reef{PF} localizes onto the $M=0$ winding sector, yielding
\beqa
\lim_{\vp \rightarrow \infty}Z_g(\vp,\bphi,\tau) = Z_g'(\tau) e^{2(g-1)\bphi}\det({\rm Im}\,\tau) \sum_{K} e^{- \pi {\rm Im}\,\tau (K^2e^{-\vp})}. \labell{PF1}
\eeqa
Consequently, the resulting \((D-1)\)-dimensional effective action is not invariant under the Buscher rules \reef{backg}. 
In this decompactification limit (large radius), the \((D-1)\)-dimensional partition function approaches the parent \(D\)-dimensional one. This is because in this limit, the KK modes become very closely spaced, and the discrete KK momentum sum effectively becomes a continuous integral.

Under T-duality  \reef{backg}, the partition function \reef{PF1} transforms to the following:
\beqa
\lim_{\vp \rightarrow -\infty}Z_g(\vp,\bphi,\tau) = Z_g'(\tau)e^{2(g-1)\bphi} \det({\rm Im}\,\tau) \sum_{ M} e^{ - \pi {\rm Im}\,\tau ( M^2 e^{\vp})}. \labell{PF2}
\eeqa
This partition function is purely stringy and cannot be obtained from the KK reduction of the \(D\)-dimensional partition function.
The corresponding \((D-1)\)-dimensional effective action is individually non-invariant under the Buscher rules. However, they transform into one another as:
\beqa
\lim_{\vp \rightarrow \infty}2\pi\int d^{D-1}x \,  \sqrt{-\bg} \, \Lambda_g(\vp , \bphi) \rightarrow\lim_{\vp \rightarrow -\infty} 2\pi\int d^{D-1}x \,  \sqrt{-\bg} \, \Lambda_g(\vp, \bphi).
\labell{cosm1}
\eeqa
    This transformation forms a $\mathbb{Z}_2$ group because applying T-duality twice returns the right-hand side to the original left-hand side.     Notably, while both the right-hand side and the left-hand side serve as a valid effective action for $|\vp|\rightarrow \infty $, they are not the complete effective action. The full action \reef{cosm}, which should be calculated directly from string theory rather than using KK reduction and T-duality, is valid for arbitrary  $\vp$.

The preceding discussion on the partition function can be extended to the calculation of S-matrix elements at the one-loop level in string theory, which are typically computed in Minkowski spacetime. For spacetimes possessing a single Killing isometry (along a coordinate \( y \)), the S-matrix method remains identical to the Minkowski case, provided the external states (vertex operators) are independent of \( y \).
In such configurations, global compactification effects are encoded in the factor \( F_2(\vp,\tau) \) \cite{Green:1982sw}. This factor effectively replaces a flat spatial direction with a circular dimension of radius \( R=e^{\vp/2} \), analogous to its role in the partition function \reef{PF}. Incorporating the radius factor \( e^{\vp/2} \) (from the measure reduction) into \( F_2(\vp, \tau) \) ensures that the product \( e^{\vp/2} F_2(\vp, \tau) \), and hence the S-matrix, is invariant under the Buscher rules.
Consequently, the effective action that reproduces these string theory S-matrix elements must also be invariant under T-duality.
In this context, one also expects that for a large radius—where winding modes do not contribute—the couplings in the effective action can be obtained from the KK reduction of the parent Minkowski spacetime effective action. The T-duality map then transforms these couplings from the large-radius regime to the small-radius regime, as demonstrated in \reef{cosm1} for the cosmological constant.

Therefore, if one calculates the loop-level \((D-1)\)-dimensional effective action directly from string theory—which naturally includes both KK and winding loop momenta—the resulting action is invariant under T-duality. In contrast, if one instead uses the KK reduction of a \(D\)-dimensional loop-level effective action (which contains no winding modes) to derive the \((D-1)\)-dimensional action, the result is not T-duality invariant. The Buscher rules then map this action to a purely stringy \((D-1)\)-dimensional effective action that cannot be derived from any parent \(D\)-dimensional effective action. 

It is worth noting that at the tree level, there are no internal momenta. Consequently, the partition function and S-matrix elements in string theory involve neither KK nor winding modes in the circle direction. It follows that the KK reduction of the $D$-dimensional effective action maps onto itself under T-duality. This results in a $D$-dimensional classical effective action that is T-duality invariant \cite{Sen:1991zi,Hohm:2014sxa}.
This situation contrasts sharply with the loop-level case. The $D$-dimensional loop-level effective action, which involves continuous internal momentum and lacks winding modes, is not T-duality invariant. In contrast, the $(D-1)$-dimensional effective action, which incorporates both KK and winding momenta, is invariant.
    The portion of the $(D-1)$-dimensional loop-level effective action obtained directly from the KK reduction of the $D$-dimensional action is only covariant under this transformation, as shown in \reef{cosm1} for the cosmological constant. 
    
In superstring and heterotic string theories, the cosmological constant vanishes perturbatively due to the exact cancellation between bosonic and fermionic contributions—a manifestation of spacetime supersymmetry. In the next section, we use the T-duality prescription above to study the one-loop effective action in the heterotic theory.
 
\section{T-duality in heterotic string theory}

    In heterotic string theory, it has been demonstrated \cite{Ellis:1987dc,Ellis:1989fi} that, for kinematic reasons related to the S-matrix elements, the one-, two-, and three-point functions vanish at one-loop order and higher genera. This led to the conclusion that all loop-level NS-NS couplings at the two-, four-, and six-derivative orders are zero in the critical 10-dimensional spacetime. Since these couplings vanish due to the S-matrix kinematics, one concludes that even after including global compactification effects—encoded in the factor \( F_2(\vp,\tau) \) \cite{Green:1982sw}—the one-, two-, and three-point functions remain zero. Consequently, all  loop-level NS-NS couplings at the two-, four-, and six-derivative orders are also zero in the 9-dimensional base space.     On the other hand, both the S-matrix method \cite{Gross:1986iv,Gross:1986mw} and T-duality \cite{Kaloper:1997ux,Garousi:2019wgz,Garousi:2019mca} require these couplings to be non-zero at the classical level. Hence, the overall dilaton dependence of these couplings, which is given by $e^{-2\Phi}$, is exact. All other classical couplings are not exact and receive quantum corrections.

The one-loop four-graviton amplitude reveals that the one-loop gravitational couplings are identical in form to the tree-level couplings, with the coupling constants subject to renormalization \cite{Ellis:1987dc}. The tree-level action is given by:
\beqa
\bS^{\rm cl.}&\supset&\int d^{10}x\sqrt{-G}e^{-2\Phi}\Big[c_1(t_8t_8+\frac{1}{4}\epsilon_8\epsilon_8)R^4+c_2\,t_8\Tr(R^2)\Tr(R^2)\Big]\,,
\eeqa
    where $c_1$ and $c_2$ are two constants, $\epsilon_8$ is the Levi-Civita tensor, and $t_8$ is a tensor that specifies how the indices of the Riemann tensor are contracted \cite{Gross:1986mw}. The $\epsilon_8\epsilon_8 R^4$ term vanishes for a four-graviton amplitude but is required by arguments from the sigma-model \cite{Grisaru:1986kw,Grisaru:1986vi} and T-duality \cite{Garousi:2020gio}. The consistency of the last term with T-duality requires its coefficient to be related to the non-zero four- and six-derivative couplings in the classical effective action \cite{Razaghian:2018svg}. Finally, the KK reduction of the classical measure $\sqrt{-G}e^{-2\Phi}$ is invariant under T-duality \reef{backg}.

The one-loop couplings have the same functional form as the tree-level action, but with different constants \( c'_1, c'_2 \) \cite{Ellis:1987dc,Abe:1988cq,Abe:1987ud}, i.e.,
 \beqa
 \bS &\supset &\int d^{10}x\sqrt{-G} \Big[ c'_1 \left(t_8t_8+\frac{1}{4}\epsilon_8\epsilon_8\right) R^4 + c'_2\, t_8\Tr(R^2)\Tr(R^2) \Big].
 \eeqa
 The KK reduction of this action is not invariant under T-duality. The reasons for this are twofold: the KK reduction of the one-loop measure \( \sqrt{-G} \) is not invariant under the transformation in \reef{backg}, and the requisite four- and six-derivative couplings are absent at loop level.
 However, as discussed in the previous section, the T-duality-invariant 9-dimensional action for an arbitrary radius is not produced by the KK reduction of the 10-dimensional action alone. The KK reduction only reproduces the correct 9-dimensional couplings in the large-radius limit (\( \vp \rightarrow\infty \)).

 If we use the reduction ansatz \reef{red}, where $G_{a y}=0$, and assume a constant radius (i.e., constant $\vp$), then the KK reduction of the above action yields:
 \beqa
 S&\supset&2\pi\int d^{9}x\sqrt{-\bg}\,e^{\vp/2}\Big[c'_1(t_8t_8+\frac{1}{4}\epsilon_8\epsilon_8)\bar{R}^4+c'_2\,t_8\Tr(\bar{R}^2)\Tr(\bar{R}^2)\Big]\,,
 \eeqa
 where $\bar{R}$ is the Riemann curvature constructed from the base space metric $\bg_{\mu\nu}$. This 9-dimensional action is a valid effective action only in the large-radius limit ($\vp\rightarrow\infty$). Under the T-duality transformation \reef{backg}, it maps to:
 \beqa
 2\pi\int d^{9}x\sqrt{-\bg}\,e^{-\vp/2}\Big[c'_1(t_8t_8+\frac{1}{4}\epsilon_8\epsilon_8)\bar{R}^4+c'_2\,t_8\Tr(\bar{R}^2)\Tr(\bar{R}^2)\Big]\,.
 \eeqa
 This is a valid effective action only in the small-radius limit ($\vp\rightarrow -\infty$). 
 
    The one-loop four-graviton S-matrix element has the kinematic factor $t_8t_8R^4$ multiplied by an integral over the modulus of the torus \cite{Schwarz:1982jn}. For an arbitrary radius, the global compactification factor \( F_2(\vp,\tau) \) should not affect the kinematic part of the four-graviton S-matrix element. Therefore, one expects the 9-dimensional effective action to have the following T-duality invariant form:
 \beqa
 S&\supset&2\pi\int d^{9}x\sqrt{-\bg}\,v(\vp)\Big[c'_1(t_8t_8+\frac{1}{4}\epsilon_8\epsilon_8)\bar{R}^4+c'_2\,t_8\Tr(\bar{R}^2)\Tr(\bar{R}^2)\Big]\,,
 \eeqa
where the function  $v(\vp)$ satisfies the relation $v(-\vp)=v(\vp)$ and has the asymptotic expansion: 
 \beqa
\lim_{\vp \rightarrow \infty} v(\vp)&=&e^{\vp/2}+\cdots\nn\\
\lim_{\vp \rightarrow -\infty} v(\vp)&=&e^{-\vp/2}+\cdots\,.
 \eeqa
The ellipses denote potential contact terms arising from S-matrix elements with both KK and winding modes in the loop. While these elements necessarily involve such states, they do not always generate contact terms; explicit calculation is required to determine the coefficients. Furthermore, one could extend this analysis to a non-constant compactification radius and a non-zero off-diagonal metric \( G_{a y} \) to incorporate their contributions into the T-duality invariant action. Such generalizations, however, lie beyond the scope of this work.

    T-duality constraints may exclude certain couplings at the classical level while permitting them in the loop-level effective action. This distinction arises from the different roles of dimensional reduction at different orders in perturbation theory.
At the classical level, the KK reduction of a $D$-dimensional coupling correctly produces all $(D-1)$-dimensional couplings that collectively satisfy the T-duality constraint. In contrast, at loop level, the KK reduction of $D$-dimensional loop-level couplings does not, by itself, generate the complete set of $(D-1)$-dimensional couplings required for T-duality invariance. In fact, the base space contains couplings that do not originate from the KK reduction of the higher-dimensional effective action.
A key example is provided by the heterotic string theory. The classical effective action lacks single-trace couplings involving more than two Yang-Mills field strengths \cite{Garousi:2024avb}, such as $\Tr(F^4)$. However, such couplings do appear in the  heterotic string theory at the one-loop level \cite{Narain:1986am,Ellis:1987dc}.

To demonstrate how these results are consistent with T-duality, consider the KK reduction of the $\Tr(F^4)$ coupling, as given in \cite{Garousi:2024avb}:
\beqa
\Tr(F_\alpha{}^\gamma F_\beta{}^\delta F_{\gamma\delta} F^{\alpha\beta})&=&\Tr(\bar{F}_a{}^c \bar{F}_b{}^d \bar{F}_{cd }\bar{F}^{ab})+4e^{\vp/2}\Tr(\bar{F}_a{}^c \bar{F}_b{}^d \bar{F}_{cd }\,{\alpha})V^{ab}+\cdots\,.
\eeqa
Here, the matrix ${\alpha}^{ij}$ is a scalar field originating from the component of the gauge field $A_\mu{}^{ij}$ along the compact $y$-direction, and $V_{ab}=\prt_a g_b-\prt_b g_a$ is the field strength associated with the off-diagonal metric component $G_{a y}=e^{\vp}g_a$. The ellipsis represents terms nonlinear in the scalar field $\alpha$, which are not of interest for our present discussion.
Under the linearized Buscher rules \cite{Garousi:2024avb}, which is given by \reef{backg} and 
\beqa
g_a'=b_a\,,\,\,\, b_a'=g_a\,,\,\,\, \alpha'^{ij}=-\alpha^{ij}\,,\,\,\, \bar{A'}_a{}^{ij}=\bar{A}_a{}^{ij}\,,
\eeqa
the right-hand side transforms as follows:
\beqa
\Tr(\bar{F}_a{}^c \bar{F}_b{}^d \bar{F}_{cd }\bar{F}^{ab})-4e^{-\vp/2}\Tr(\bar{F}_a{}^c \bar{F}_b{}^d \bar{F}_{cd }\,{\alpha})W^{ab}\,,\labell{TF4}
\eeqa
where $W_{ab}$ is the field strength of the off-diagonal Kalb-Ramond field $B_{a y}=b_a$.
 Crucially, there is no 10-dimensional coupling whose KK reduction produces the 9-dimensional couplings in \reef{TF4}. We therefore conclude that T-duality forbids the $\Tr(F^4)$ coupling from appearing in the classical effective action. The absence of a single-trace term is also consistent with the cosmological reduction of the classical couplings possessing $O(9,25)$-symmetry \cite{Garousi:2024wlw}.

However, T-duality permits this coupling at the one-loop level. Consider the KK reduction of the one-loop coupling:
\beqa
\int d^{10}x\sqrt{-G}\,\Tr(F^4)=2\pi\int d^9 x \sqrt{-\bar{g}}e^{\vp/2}\Big[\Tr(\bar{F}_a{}^c \bar{F}_b{}^d \bar{F}_{cd }\bar{F}^{ab})+4e^{\vp/2}\Tr(\bar{F}_a{}^c \bar{F}_b{}^d \bar{F}_{cd }\,{\alpha})V^{ab}\Big]\,.
\eeqa
This expression yields the correct couplings in the 9-dimensional base space only in the large-radius limit ($\vp\rightarrow\infty$). In this regime, it should be reproduced by a torus-level S-matrix element involving only KK momentum circulating in the loop.
Under T-duality, the right-hand side transforms to:
\beqa
2\pi\int d^9 x \sqrt{-\bar{g}}e^{-\vp/2}\Big[\Tr(\bar{F}_a{}^c \bar{F}_b{}^d \bar{F}_{cd }\bar{F}^{ab})-4e^{-\vp/2}\Tr(\bar{F}_a{}^c \bar{F}_b{}^d \bar{F}_{cd }\,{\alpha})W^{ab}\Big]\,.
\eeqa
This represents the correct couplings in the 9-dimensional base space only in the small-radius limit ($\vp\rightarrow-\infty$). These couplings should be reproduced by a torus-level S-matrix element involving only winding momentum circulating in the loop.
For an arbitrary radius $\vp$, one must include the contributions from both KK and winding momenta circulating around the torus, which is a non-trivial task. In this general case, the resulting 9-dimensional effective action must be invariant under the Buscher rules. This requires the action to take the form:
\beqa
S&\supset&2\pi\int d^9 x \sqrt{-\bar{g}}\Big[f(\vp)\Tr(\bar{F}_a{}^c \bar{F}_b{}^d \bar{F}_{cd }\bar{F}^{ab})\nn\\&&+4g(\vp)\Tr(\bar{F}_a{}^c \bar{F}_b{}^d \bar{F}_{cd }\,{\alpha})V^{ab}-4g(-\vp)\Tr(\bar{F}_a{}^c \bar{F}_b{}^d \bar{F}_{cd }\,{\alpha})W^{ab}\Big]\,,
\eeqa
 where the function $f(\vp)$ satisfies the relation $f(-\vp)=f(\vp)$, and  where $f(\vp)$ and $g(\vp)$ have the following  asymptotic expansions: 
 \beqa
 \lim_{\vp \rightarrow \infty} f(\vp)=e^{\vp/2}+\cdots\,, &&\lim_{\vp \rightarrow -\infty} f(\vp)=e^{-\vp/2}+\cdots\,,\nn\\
\lim_{\vp \rightarrow \infty}g(\vp)=e^{\vp}+\cdots\,,&&\lim_{\vp \rightarrow -\infty}g(\vp)=\cdots\,.
 \eeqa
The ellipses denote potential contact terms arising from S-matrix elements with both KK and winding modes in the loop. Although T-duality alone does not fix these couplings, they may be constrained by other symmetries of the theory. For instance, in type II theory—which possesses both T-duality and S-duality—the corresponding moduli functions could be determined by the constraints of S-duality. We will explore the T-duality and S-duality of type II theory in the following section.

 \section{T-duality in type II string theory}
 
In type II superstring theories, no couplings exist at the four- and six-derivative levels at any genus in the critical dimension. Furthermore, the dilaton dependence of the classical two-derivative action is exact. The classical effective action for these theories includes the following gravitational couplings at the eight-derivative order \cite{Gross:1986iv,Gross:1986mw,Grisaru:1986kw,Grisaru:1986vi}:
 \beqa
\bS^{\rm cl.}&\supset&a_1\int d^{10}x\sqrt{-G}e^{-2\Phi}\Big[(t_8t_8+\frac{1}{4}\epsilon_8\epsilon_8)R^4\Big]\,,
\eeqa
where $a_1$ is a constant proportional to $\z(3)$. The consistency of these couplings with T-duality uniquely determines all NS-NS couplings at this order in the classical effective action \cite{Garousi:2020gio}. The difference between the type IIA and type IIB classical actions lies solely in the RR sector, which is not of interest in this context.

At the one-loop level, however, the NS-NS couplings in the two theories are not identical. The gravitational couplings in type IIA theory differ from their classical counterparts. Furthermore, type IIA theory includes Chern-Simons couplings that depend linearly on the B-field. The effective action is  \cite{Green:1984sg,Sakai:1986bi,Russo:1997mk,Vafa:1995fj,Antoniadis:1997eg}
\beqa
\bS^{\rm IIA}&\supset&a_1'\int d^{10}x\sqrt{-G}\Big[(t_8t_8-\frac{1}{4}\epsilon_8\epsilon_8)R^4-\frac{1}{2}\epsilon_{10}t_8 B R^4\Big]\,,\labell{IIA}
\eeqa
where \(a_1'\) is a constant proportional to \(\pi^2\).

Type IIB theory, on the other hand, has gravitational couplings similar to its classical counterpart. This is consistent with the S-duality of type IIB string theory, which requires the loop-level couplings to be related to the classical ones by extending the overall dilaton factor to be S-duality invariant \cite{Green:1997tv}. The gravitational couplings are \cite{Sakai:1986bi,Russo:1997mk,Vafa:1995fj,Antoniadis:1997eg}:
\beqa
\bS^{\rm IIB}&\supset&a_1'\int d^{10}x\sqrt{-G}\Big[(t_8t_8+\frac{1}{4}\epsilon_8\epsilon_8)R^4\Big]\,.\labell{IIB}
\eeqa
The absence of a Chern-Simons coupling in the type IIB effective action at the one-loop level is consistent with S-duality, as no such coupling exists in the classical effective action.

Under the KK reduction with ansatz \reef{red} and a constant radius, the 10-dimensional type IIA effective action \reef{IIA} becomes:
\beqa
S^{\rm IIA}&\supset&2\pi a_1'\int d^{9}x\sqrt{-\bg}\Big[e^{\vp/2}(t_8t_8-\frac{1}{4}\epsilon_8\epsilon_8)\bar{R}^4-\epsilon_{9}t_8 b \bar{R}^4\Big]\,.\labell{IIA}
\eeqa
Note that since the Chern-Simons term is topological, its reduction does not depend on the radius of the compact space. The above action is the correct 9-dimensional effective action of type IIA theory for a large radius ($\vp\rightarrow\infty$), where only the KK momentum circulates in the loop. Under T-duality, it transforms into the following couplings in type IIB theory:
\beqa
S^{\rm IIB}&\supset&2\pi a_1'\int d^{9}x\sqrt{-\bg}\Big[e^{-\vp/2}(t_8t_8-\frac{1}{4}\epsilon_8\epsilon_8)\bar{R}^4-\epsilon_{9}t_8 g \bar{R}^4\Big]\,.\labell{IIB}
\eeqa
These couplings are the correct 9-dimensional couplings in type IIB theory for a small radius ($\vp\rightarrow-\infty$), where only the winding momentum circulates in the loop. It is important to note that since the vector \(g_a\) is related to  the reduction of the 10-dimensional metric via \(G_{a y} = e^{\vp} g_a\),  the Chern-Simons term cannot be generated from the KK reduction of any covariant and gauge-invariant 10-dimensional coupling in type IIB theory.

On the other hand, the KK reduction of the 10-dimensional type IIB effective action \reef{IIB} becomes:
\beqa
S^{\rm IIB}&\supset&2\pi a_1'\int d^{9}x\sqrt{-\bg}\,e^{\vp/2}\Big[(t_8t_8+\frac{1}{4}\epsilon_8\epsilon_8)\bar{R}^4\Big]\,.
\eeqa
This is the correct 9-dimensional effective action of type IIB theory for a large radius ($\vp\rightarrow\infty$), where only the KK momentum circulates in the loop. Under T-duality, it transforms into the following couplings in type IIA theory:
\beqa
S^{\rm IIA}&\supset&2\pi a_1'\int d^{9}x\sqrt{-\bg}\,e^{-\vp/2}\Big[(t_8t_8+\frac{1}{4}\epsilon_8\epsilon_8)\bar{R}^4\Big]\,.
\eeqa
These couplings are the correct 9-dimensional couplings in type IIA theory for a small radius ($\vp\rightarrow-\infty$), where only the winding momentum circulates in the loop.

The Chern-Simons terms in \reef{IIA} and \reef{IIB} are topological; hence, we do not expect their coefficients to depend on the radius. 
By combining the large-radius and small-radius limits of the above couplings, one can infer that the eight-derivative effective action for an arbitrary radius should take the following form:
\beqa
S^{\rm IIA}&\supset&2\pi a_1'\int d^{9}x\sqrt{-\bg}\Big[h(\vp)t_8t_8\bar{R}^4-\frac{k(\vp)}{4}\epsilon_8\epsilon_8\bar{R}^4-\epsilon_{9}t_8 b \bar{R}^4\Big],\nn\\
S^{\rm IIB}&\supset&2\pi a_1'\int d^{9}x\sqrt{-\bg}\Big[h(\vp)t_8t_8\bar{R}^4+\frac{k(\vp)}{4}\epsilon_8\epsilon_8\bar{R}^4-\epsilon_{9}t_8 g \bar{R}^4\Big],\labell{FAB}
\eeqa
where the functions  $h(\vp)$ and  $k(\vp)$ satisfy the relations $h(-\vp)=h(\vp)$ and $k(-\vp)=-k(\vp)$. These functions have the following asymptotic expansions:
\beqa
 \lim_{\vp \rightarrow \infty} h(\vp)=e^{\vp/2}+\cdots\,, &&\lim_{\vp \rightarrow -\infty} h(\vp)=e^{-\vp/2}+\cdots\,,\nn\\
\lim_{\vp \rightarrow \infty}k(\vp)=e^{\vp/2}+\cdots\,,&&\lim_{\vp \rightarrow -\infty}k(\vp)=-e^{-\vp/2}+\cdots\,.\labell{dots}
 \eeqa
 The ellipses denote potential contact terms arising from S-matrix elements with both KK and winding modes in the loop. The two sets of couplings in \reef{FAB} are related by T-duality. While the explicit forms of the functions \(h(\vp)\) and \(k(\vp)\) are expected from direct calculation, they can be constrained by symmetry. Specifically, \(h(\vp)\) can be determined by incorporating the finite-radius correction factor \(F_2(\vp, \tau)\) into the four-graviton S-matrix elements \cite{Green:1982sw} and integrating over the modulus \(\tau\). However, since these functions are common to both type IIA and type IIB theories, and type IIB enjoys S-duality, they are ultimately fixed by the S-duality constraints of the type IIB theory. We will determine these functions using S-duality in the next subsection.


\subsection{Constraining moduli functions with S-duality}

Unlike T-duality transformations, which are given in the string frame, the S-duality transformations of type IIB theory are properly defined in the Einstein frame, where the metric is related by \(G_{\mu\nu}=e^{\Phi/2}G^E_{\mu\nu}\). Applying this transformation to the circularly reduced metric in \reef{red}, one finds the following relations between the string frame and Einstein frame base space fields:
\beqa
\bg_{ab} = e^{\Phi/2}\bg_{ab}^E\,,\quad  \vp = \frac{1}{2}\Phi + \vp^E.\labell{Evp}
\eeqa
Under S-duality, the ten-dimensional Einstein frame metric is invariant; hence,
\beqa
\bg_{ab}^E\rightarrow \bg_{ab}^E\,,\quad \vp^E\rightarrow \vp^E\,.\labell{ggp}
\eeqa

The couplings in type IIB theory are expected to be invariant under S-duality for any value of the radion \(\vp\). The first two terms in \reef{FAB} have the following structure:
\beqa
\sqrt{-\bg}\Big[h(\vp)t_8t_8\bR^4+\frac{k(\vp)}{4}\epsilon_8\epsilon_8\bR^4\Big].
\eeqa
In the Einstein frame and for a constant dilaton, this becomes
\beqa
\sqrt{-\bg^E}e^{\Phi/4}\Big[h(\Phi/2+\vp^E)t_8t_8(\bR^E)^4+\frac{k(\Phi/2+\vp^E)}{4}\epsilon_8\epsilon_8(\bR^E)^4\Big]\labell{Ef}
\eeqa
where \(\bR^E\) is the Riemann curvature constructed from \(\bg_{ab}^E\).

In the large-radius limit, it becomes
\beqa
\sqrt{-\bg^E}e^{\Phi/2}\Big[e^{\vp^E/2}t_8t_8(\bR^E)^4+\frac{1}{4}e^{\vp^E/2}\epsilon_8\epsilon_8(\bR^E)^4\Big].
\eeqa
There is a similar term with an overall dilaton factor \(e^{-3\Phi/2}\) in the tree-level effective action. These two perturbative contributions, along with some non-perturbative contributions, combine to be invariant under the S-duality group \(SL(2,\mathbb{Z})\)\cite{Green:1997tv}.

Given that the tree-level effective action and the non-perturbative contributions are all included in the above action to achieve invariance, one concludes that the effective action \reef{Ef} must be invariant for all other parts of the functions \(h(\vp)\) and \(k(\vp)\) by themselves. This uniquely fixes these functions, allowing only one other contribution, which corresponds to the small-radius limit. Hence, S-duality requires the ellipses in \reef{dots} to be zero.  The last term in \reef{FAB} is also invariant under S-duality for a constant dilaton. This invariance holds because the vector \(g_a\) remains unchanged between the string and Einstein frames and is S-duality invariant. Furthermore, the term \(\epsilon'_{9}t_8 g \bar{R}^4\)—where \(\epsilon'_{9}\) is the Levi-Civita symbol—contains four Riemann curvatures and four inverse metrics, ensuring the correct transformation behavior.

\section{Discussion}

In this work, we examine the background dependence of the string effective action across different genus orders. Our findings indicate that the classical (\(g=0\)) effective action is background-independent, a direct consequence of the fact that sphere-level S-matrix elements—from which the action is derived—involve no internal momentum integration. In contrast, quantum corrections at higher genus (\(g>0\)) introduce significant background dependence.
This dependence arises because loop-level S-matrix elements involve internal momentum circulating in the loops. In a Minkowski spacetime, this momentum is continuous in all directions. However, when spatial dimensions are compactified on circles, the internal momentum along those directions becomes discrete and includes both KK and winding modes. While the KK momentum is simply the discretization of the continuous momentum, winding momentum is a purely string-theoretic feature absent in a non-compact Minkowski background. Consequently, the resulting S-matrix elements, and the effective actions derived from them, differ fundamentally between a fully non-compact spacetime and one with compactified dimensions.
We investigate this background dependence by studying the effective action for spacetimes with one Killing isometry, using T-duality invariance as our primary theoretical constraint.

At the classical level,  the T-duality completely determines the structure of both the $D$-dimensional effective action and the $(D-1)$-dimensional base space effective action for spacetimes with a Killing circle. This fixes all but a few parameters, which can subsequently be determined through sphere-level S-matrix calculations in flat spacetime. The resulting effective action is valid for spacetimes without a Killing circle, demonstrating its background independence.
However, the situation changes dramatically at higher genus. Background dependence makes it generally impossible to determine the $D$-dimensional effective action using T-duality alone. In fact, T-duality cannot even fully constrain the form of the $(D-1)$-dimensional base space effective action when one spatial dimension is a Killing circle.

In this paper, we propose that T-duality may still relate the $(D-1)$-dimensional base space effective action to the $D$-dimensional effective action in two specific regimes: when the circle radius is very large and when it is very small. Our prescription is as follows: in the large-radius limit, winding momentum in loop-level S-matrix elements can be neglected. The resulting S-matrix elements then match those of a loop-level calculation in a non-compact flat spacetime, which has no winding modes. Consequently, the corresponding $(D-1)$-dimensional couplings can be found from the KK reduction of the $D$-dimensional effective action. By applying a T-duality transformation to this result, one can then derive the $(D-1)$-dimensional effective action for the very small-radius limit. This small-radius action, which corresponds to S-matrix elements where only winding momentum circulates in the loops, constitutes a purely stringy effect. It cannot be obtained from the KK reduction of any covariant and gauge-invariant $D$-dimensional coupling.
Using this prescription, we derive several 9-dimensional one-loop couplings in heterotic and type II superstring theories for both large- and small-radius limits. In the type II case, we show that the S-duality of the effective action requires the moduli-dependence to be exclusively restricted to these large- and small-radius limits.

The background independence of the classical effective action implies that the $D$-dimensional and $(D-1)$-dimensional effective actions are equivalent; that is, the latter can be derived from the KK reduction of the former when one spatial dimension is compactified on a circle. In contrast, the background dependence of the loop-level effective action means this equivalence no longer holds: the $(D-1)$-dimensional and $D$-dimensional actions are fundamentally different.
A parallel logic applies to D-brane effective actions. The world-volume couplings involving only massless open string fields at the disk level are classical, and T-duality can be used to determine them \cite{Garousi:2022rcv}. In this case, background independence implicitly assumes that the D-brane couplings are independent of the brane's dimension. However, this assumption fails for couplings involving both massless open and closed string fields, which represent genuine quantum effects. For these, the T-duality constraint is violated if one assumes identical world-volume couplings for a D$_p$-brane and a D$_{p-1}$-brane \cite{Garousi:2024pqc}.


\end{document}